\documentstyle[times,aps,pre,eqsecnum,epsfig,balanced]{revtex}
\tighten

\newcommand{\BE}{\begin{equation}}
\newcommand{\EE}{\end{equation}}
\newcommand{\BA}{\begin{eqnarray}}
\newcommand{\EA}{\end{eqnarray}}
\newcommand{\bc}{\begin{center}}
\newcommand{\ec}{\end{center}}

\begin{document}

\draft

\title{Frozen spatial chaos induced by boundaries}

\author{
V\'\i ctor M. Egu\'\i luz\cite{email}, Emilio Hern\'andez-Garc\'\i a,
Oreste Piro, and Salvador Balle}
\address{
Instituto Mediterr\'aneo de Estudios Avanzados IMEDEA\cite{imedea} 
(CSIC-UIB)\\ E-07071 Palma de Mallorca (Spain) 
} 

\date{\today}

\maketitle

\begin{abstract}

We show that rather simple but non-trivial boundary conditions 
could induce the appearance of {\sl spatial chaos} (that is 
stationary, stable, but spatially disordered configurations) in 
extended dynamical systems with very simple dynamics. We exemplify 
the phenomenon with a nonlinear reaction-diffusion equation in a 
two-dimensional undulated domain. Concepts from the theory of 
dynamical systems, and a transverse-single-mode approximation are 
used to describe the spatially chaotic structures. 

\end{abstract}


\begin{twocolumns}
\section{INTRODUCTION}
\label{sec:Intro}

In the past few decades, considerable understanding of the 
phenomenon of temporal chaos in dynamical systems of few degrees 
of freedom has been achieved \cite{Hao90,Ott,Yorke}. On the other 
hand, spatio-temporal chaos in extended dynamical systems with 
infinitely many degrees of freedom is currently under very active 
investigation\cite{Cross94,Cross93}. It is 
remarkable however, that an area of problems laying somehow 
between the two extremes has not received so much attention, 
namely, purely spatial chaos as a stationary attractor of extended 
dynamical systems 
\cite{Coullet87,Rabinovich92,montagne96,Chang94,Balmforth96}. 

The possible existence of this kind of attractors was first 
suggested by Ruelle \cite{Ruelle83} in the context of equilibrium 
phases. He pointed out the parallelism between a time-dependent 
differentiable dynamical system and the space dependence of 
equilibrium states in statistical mechanics. He then raised the 
question as to whether the existence of turbulent crystals could 
be the natural next step towards complexity after spatially 
homogeneous, periodic, and quasi-periodic equilibrium phases have 
been found. Newell and Pomeau \cite{Newell93} gave some 
conditions under which such a turbulent crystal would exist in 
pattern-forming systems described by a free energy. Theoretical 
and experimental work on modulated phases and 
commensurate-incommensurate transitions \cite{Bak82} represent 
additional concrete results along these lines. 

In the context of fluid dynamics, the existence of spatially chaotic,
but temporally steady solutions would also fill a conceptual gap
between two well-studied complex phenomena: Lagrangian chaos, and
Eulerian chaos or turbulence. The former refers to the chaotic motion
of a fluid parcel which might occur even in laminar and, in three
dimensions, steady flows \cite{Jones89,LeGuer,Julyan97}. On the
other extreme, the road to turbulence is usually associated to a
hierarchy of instabilities leading to increasingly spatio-temporally
chaotic Eulerian velocity
fields. Frozen spatial chaos would then refer in this context to a
third possibility: a stationary flow spatially chaotic in the Euler 
description. 

By now, many extended dynamical systems displaying spatial chaos have
been identified. Most of the studies are concerned with 
one-dimensionally extended systems. They are specially suitable to 
analysis because their steady state configurations depend just on the
unique spatial coordinate. These configurations are solutions of sets
of ordinary differential equations (the {\sl spatial} dynamical system)
with the space variable as the independent variable. The standard
theory of low-dimensional dynamical systems can be used to describe
such configurations, by just considering the spatial coordinate as a
fictitious time. Rigidly traveling waves with spatial chaotic structure
can also be considered as a case of spatial chaos in a moving reference
frame \cite{montagne96,Chang94}. 

Spatial chaos may appear when the {\sl spatial} dynamical system has a
sufficiently high dimensional phase space. This high dimensionality
may arise from either a) the presence of high-order spatial
derivatives in a single evolution equation as in the cases of the
Swift-Hohenberg equation \cite{EHG92,Gorshkov94}, and
Kuramoto-Sivashinsky and related models \cite{Chang94,Balmforth96}, b)
the coupling of several fields each one satisfying a lower order
differential equation as in (the real and imaginary parts of) the
complex Ginzburg-Landau equation\cite{montagne96} which supports
chaotic travelling waves, or
c) explicit space dependent forcing terms as in \cite{Coullet87} or
\cite{Malkov95} . 

Consideration of two-dimensional spatial chaos has been much 
scarce. In absence of a simple connection with conventional 
dynamical systems theory, the very concept of chaos in the spatial 
configuration should be properly defined for the general 
two-dimensional case. A rather complete formalism generalizing 
dynamical system tools (entropies, dimensions,...) to 
multidimensional spatial chaos has been 
developed\cite{Rabinovich92,Afraimovich92}, and some examples 
examined \cite{Rabinovich92,Gorshkov94}. On the other hand, as the 
number of dimensions increases, a much larger variety of 
non-trivial boundary condition classes surely leads to a greater 
richness in the expected properties of the steady field 
configurations. A well-posed question is then whether relatively 
simple boundary conditions may lead to steady spatially chaotic 
configurations. The main purpose of this Paper is to address this 
question. 

In addition to the existence of chaotic spatial configurations, it is
important to study also their stability in time. A stationary state
will only be physically observable if it is stable or at least
long-lived. It turns out that the temporal stability of the stationary
solutions is in general unrelated to the stability of these
configurations considered as orbits of the spatial dynamical system. In
the examples cited above, there are cases of both stable and unstable
space-chaotic configurations, but instability seems to be more
frequent. As a consequence, spatial chaos has been generally considered
of limited physical relevance. 

In this Paper we show that rather simple undulated strip-like domain
shapes can induce, in a very simple nonlinear extended dynamical
system, the formation of patterns that are both {\sl spatially
chaotic} and {\sl temporally attracting}. The kind of {\sl modulated
boundaries} we use could be easily implemented in standard
experimental pattern-formation set-up's such as Faraday waves,
convection cells, or open flows. In fact, our work was originally
motivated by the observation, in a fluid dynamics experimental setup
consisting of a periodic array of pipe bends, that the transverse
profile of the steady flow does not necessarily repeat itself with the
same periodicity of the array \cite{LeGuer}. 

In Section~\ref{Dynsys} we present the particular model that we study 
and perform a preliminary analysis of its behavior. In
Section~\ref{Low} a single-transverse-mode approximation is introduced
and we use it to predict the existence of boundary-induced spatial
chaos. Numerical simulations are presented in Section~\ref{Num} to 
substantiate our claims beyond the validity of the previous 
approximation. Finally we summarize the results and open problems in
the Conclusions. 

\section{A REACTION-DIFFUSION EQUATION IN A STRIP-SHAPED DOMAIN}
\label{Dynsys}

As stated before, the application of the theory of dynamical systems to
the study of stationary spatial configurations of one-dimensionally
extended systems is direct. A stationary pattern satisfies, in general,
a system of ordinary differential equations with the spatial coordinate
as its independent variable which we can think of as a time. Parity
symmetry in a spatial coordinate will appear as time-reversal symmetry
after reinterpretation of this coordinate as time. 

The general study of spatial chaos in several spatial dimensions
requires the notion of translational dynamical systems with $d$ times
\cite{Rabinovich92,Afraimovich92}. There are situations, however, 
where such formalism is not necessary because the two independent
spatial directions are distinguished by the geometry of the system,
so that one of them naturally plays the role of time. In this way, the
spatial variation in one direction would be interpreted as time
evolution of a one-dimensional field that only depends on the
remaining spatial coordinate. Particularly suited to our approach will
be the case of two-dimensional extended systems in strip-shaped 
regions much longer (ideally infinite) in the time-like direction than
in the space-like one. If the strip is narrow enough, only patterns
composed of one or few transverse spatial modes will be allowed and
spatial chaos could be readily defined and identified in terms of the
usual concepts of dynamical systems theory. 

In order to concentrate on spatial chaos purely induced by 
boundary effects, we consider a very simple model equation 
containing only up to second order derivatives and a single field 
variable, a reaction-diffusion equation of the Fisher-Kolmogorov 
type: 
\begin{equation}
\partial_{t} {\bf \psi} = \nabla^{2} \psi + a \psi -\psi^{3} \ \ ,
\label{fk}
\end{equation}
with appropriate boundary conditions for the real field $\psi({\bf x},t)$. 
The real linear coefficient $a$ can be absorbed rescaling the variables, but
we find convenient to keep it explicit in the equation. Equation~(\ref{fk})
appears in several contexts including phase transitions, where it takes the 
name of real Ginzburg-Landau equation or time-dependent Ginzburg-Landau
model \cite{phaset}, and population dynamics \cite{Murray93}. The dynamics
of (\ref{fk}) can be written as purely relaxational \cite{potential} in a
functional Lyapunov potential 
$V[\psi]$: 
\BE 
\partial_t\psi = - {\delta V[\psi] \over \delta \psi} \label{relax}
\EE
with 
\BE V[\psi]=\int_D d{\bf x}\left( \frac{1}{2}|\nabla\psi|^2 - \frac{a}{2}
 \psi^2+ \frac{1}{4}\psi^4 \right)\label{V} + S[\psi]~,
\EE
where the integral is over the domain $D$. The surface term $S[\psi]$ takes 
into account the effects of boundary conditions over the domain limit, and it
vanishes when periodic, null Dirichlet ($\psi = 0$) or null Neumann
($\partial_n\psi=0$) boundary conditions are specified. It follows from
Eq.~(\ref{relax}) that $V$ can only decrease with time. The relaxational
character of Eq.~(\ref{relax}) implies also that the only asymptotic states
are fixed points. Therefore, this model does not display any limit-cycle
oscillations or more complex dynamics such as temporal chaos in any number of
spatial dimensions.

Equation~(\ref{fk}) has been extensively studied in one and two 
dimensions. In one dimension, for infinite systems, we have the 
following situations: For $a<0$, $\psi= 0$ is the only stationary 
solution and is stable under time evolution. At $a = 0$ a 
pitchfork bifurcation occurs and the former trivial solution 
looses its stability. For $a>0$ some of the stationary solutions 
are the following: 
\begin{enumerate}
\item Homogeneous solutions:
$\psi(x,t)= 0$, and $\psi (x,t)= \pm \sqrt{a} \equiv \psi_\pm $.
\item Kink type solutions: $\psi (x,t) = \pm \sqrt{a}\tanh(\sqrt{\frac{a}{2}} x)$.
\item Periodic solutions: $\psi(x,t) = \sqrt{a-k^2} \sin (k x) + ...$,
where the ellipsis stands for higher order harmonics.
\end{enumerate}
Only $\psi_\pm$ and the kinks are now linearly stable. 

For finite systems, the boundary conditions normally restrict the 
variety of solutions either by selecting some values of $k$, or by 
favoring either the kink or the $\psi_\pm$ solutions. 
Additionally, the boundary conditions may have as an effect a 
shift in the pitchfork bifurcation point $a=0$ to a different 
value $a_c\neq0$ and a change in the coefficients appearing in the 
solutions. 

The typical time evolution of initial field distributions leads to 
the formation of domains where values of $\psi$ close to either 
$\psi_+$ or $\psi_-$ dominate. These domains are separated by 
kink- or anti-kink-type walls that can move into each other 
producing a mutual annihilation. By this mechanism, small domains 
disappear and feed the larger domains whose sizes then increase 
logarithmically in time until one of the stationary solutions, 
prevailing by chance, takes over the whole system. 

In two dimensions, the dynamics typically consists of the coarsening of
domains of $\psi_+$ and $\psi_-$ phases whose typical size grows as
the square root of time\cite{phaset}. Interfaces between the two phases
are locally similar to the one-dimensional kink solutions. The gradient
term in Eq.~(\ref{V}) is important at the interfaces giving a positive
contribution to the Lyapunov potential. Since the dynamics always
minimizes $V$, it tends to reduce the length of these interfaces. This
reduction is achieved by the shrinking and ulterior collapse of the
smallest domains to contribute to the
(square-root) growth of the remaining ones. 

It was shown by Collet \cite{Collet94} in a more general context that
the time evolution and final states of Eq.~(\ref{fk}) in finite domains
are similar to those in an infinite system except in a boundary layer
around the border whose size depends on the $a$ parameter. This result
holds in both one and two dimensions. Thus, in order to observe the
influence of boundaries on pattern evolution, we need to consider a
domain small enough at least in one of the directions. In a stripped
domain, elongated in the $x$ direction, this small dimension will be
the transverse $y$ direction. The domain will be limited in this
transverse dimension by the boundaries $y_0(x)$ and $y_1(x)$, where the
function $\psi(x,y,t)$ will take values $\psi_0(x)$ y $\psi_1(x)$,
respectively (Dirichlet conditions). 

In the one-dimensional case, conventional dynamical systems theory 
implies that there are no chaotic stationary solutions to
Eq.~(\ref{fk}) because the spatial dynamical system is just a second 
order ordinary differential equation. However, chaos can arise if some
$x$-dependent periodic forcing is added to the equation. These
arguments do not apply directly to the two-dimensional case. However,
it is tempting to think of undulations of the lateral boundaries as a
kind of periodic forcing on the longitudinal coordinate. This suggests
the possibility of finding chaotic structures in the $x$-direction
induced by undulated boundaries. 

\begin{figure} 
\bc
\epsfig{file=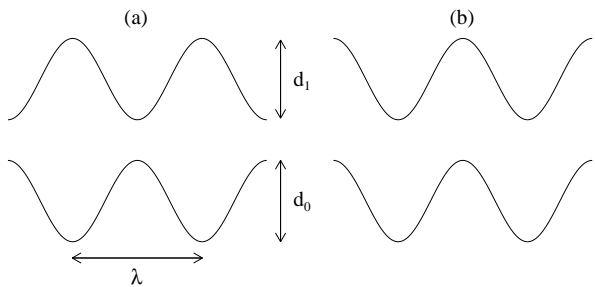,width=.5\linewidth,angle=90}
\caption{ 
Examples of stripped channels enclosed in oscillating walls:
$\alpha = 2 \pi / \lambda$; $d_0=d_1$, and (a) $\phi=\pi$,
(b) $\phi=0$.}
\label{fdomain} 
\ec
\end{figure}

As a particular case we consider domains limited by two 
sinusoidal boundaries. With applications to hydrodynamics in mind 
we think of these domains as channels with sinusoidal banks. 

\BA y_1(x) &=& \frac {d_1}{2}(1-\cos(\alpha x)) 
\label{boundary1}\\ 
y_0(x) &=& -1 -\frac {d_0}{2}(1+\cos(\alpha x 
+ \phi)) 
\label{boundary2}
\EA

Here $d_1$, $d_0$ are the amplitude of the undulation of each bank,
$\alpha$ is the spatial frequency which we assume to be the same for
both banks and $\phi$ is their mutual phase mismatch.
Figure~(\ref{fdomain}) shows a few typical shapes for our channel-like
domains. The case $d_1=d_0$ and $\phi=\pi$ gives a sausage shaped
channel with symmetrically and sinusoidally varying width. On the other
hand, $d_1=d_0$ and $\phi=0$ sets the boundaries in phase and
corresponds to a domain with the form of a sinusoidally meandering
channel of constant $y$-width.

We stress that we want to consider the simplest situation that may 
display spatial chaos. Consideration of more complex equations 
exhibiting spatial chaos even with simple boundaries, or more 
complex boundaries such as incommensurate oscillations for the 
upper and lower banks (corresponding to quasi-periodic forcing) 
would only enrich the complexity of stationary solutions. 

A complete definition of the model requires also the specification of 
boundary conditions on the longitudinal $x$ direction. The analogy 
with a temporal variable would be better for domains infinite in the
$x$ direction, with only the weak requirement of boundedness for
$\psi$. However, an infinite domain is inadequate for the numerical
approaches to be described below. In our calculations we would need to
impose periodic boundary conditions (of period $L$) along the $x$
direction. In this way we are restricting the class of solutions to
periodic orbits of period $L$ or less in the time-like coordinate. We
will still be able to identify as spatially chaotic the configurations
that have the maximal period $L$, provided this period increases and
the periodic orbit approaches a chaotic trajectory as system size $L$
increases. Subtle considerations such as Lyapunov number computations
for such limiting orbits will be addressed elsewhere. 

To perform numerical simulations we can choose between several 
strategies. If we are only interested in stationary states, we can 
numerically solve the time-independent version of Eq.~(\ref{fk}) 
by means of finite elements or finite differences. These methods 
can be implemented to find solutions that may or may not be stable 
under time evolution. Another possibility is to follow the 
dynamics of the full Eq.~(\ref{fk}) until a stationary state is 
reached. In this way, only attracting (i.e. stable) stationary 
solutions can be found (remind that only stationary 
attractors are allowed by this purely relaxational dynamics). In 
any case, a convenient way to handle the boundary conditions is to 
map the region limited by $y_0(x)$ and $y_1(x)$ (and by $x=0,L$) 
to a rectangular one: $\tilde y_1=1$, $\tilde y_0=0$, and $x=0,L$. 
For arbitrary functions $y_0(x)$ and $y_1(x)$, the map
$(x,y)\longmapsto (x,\tilde y)$, with
\BE
\tilde y = (y -y_0) / (y_1 - y_0)~,
\EE
transforms Eq.~(\ref{fk}) into an equation for
$\tilde\psi(x,\tilde y,t) \equiv \psi(x,y,t)$: 
\BA
\partial_{t} \tilde\psi =
\partial^{2}_{xx} \tilde\psi &+& {\cal F}(x)\partial^{2}_{\tilde y\tilde y}
\tilde\psi + {\cal G}(x)\partial^{2}_{x\tilde y} \tilde\psi \nonumber\\
&+& {\cal H}(x)\partial_{\tilde y} \tilde\psi + a \tilde\psi - \tilde\psi^{3}
\label{nfk1}
\EA
where
\BA
{\cal F}(x) & = & \frac {1+(\Delta_x \tilde y + y_{0x})^2}{\Delta^2}\\
\label{nfk2}
{\cal G}(x) & = & -2 \frac {\Delta_x \tilde y + y_{0x}}{\Delta}\\
\label{nfk3}
{\cal H}(x) & = & \frac {\tilde y (2\Delta_x^2 - \Delta \Delta_{xx})
- \Delta y_{0xx} +2\Delta_x y_{0x}}{\Delta^2} \\
\label{nfk4}
\Delta(x) &=& y_{1}(x)-y_{0}(x)\\
\label{delta}
\Delta_x(x) &=& \frac{d}{d x} \Delta(x)\\ 
\label{deltax}
\Delta_{xx}(x) &=& \frac{d^2}{dx^2} \Delta(x)\\ 
\label{deltaxx}
y_{0,1x} &=& \frac{d}{d x}y_{0,1} 
\label{yx}
\EA
Here $\Delta (x)$ is the transverse distance between the boundaries and
gives the width of the strip. If this width does not vary along $x$,
then $\Delta_x (x) = 0 = \Delta_{xx}(x)$.

The new transverse boundary conditions are
\BE
\tilde\psi(x,\tilde y=0)=\psi_0(x), \ \ \tilde\psi(x,\tilde
y=1)=\psi_1(x)~.
\label{bc}
\EE

A first observation is that the shape of the domain boundaries is 
reflected as a parametric forcing of the equation in the new 
coordinates. For example, for the simplest case of a meandering 
channel given by Eqs.~(\ref{boundary1}) and (\ref{boundary2}) with 
$\phi =0$, and $d_0=d_1=d$ (Fig.~\ref{fdomain}b):
\BA 
\partial_{t} \tilde\psi =
\partial^{2}_{xx} \tilde\psi &+& {\cal F}(x)\partial^{2}_{\tilde y\tilde y}
\tilde\psi + {\cal G}(x)\partial^{2}_{x\tilde y} \tilde\psi \nonumber\\
&+& {\cal H}(x)\partial_{\tilde y} \tilde\psi
+ a \tilde\psi - \tilde\psi^{3}
\label{phasefk1}
\EA
\BA
{\cal F}(x) &=& \frac {1+\left( \frac {d \alpha \sin(\alpha x)}{2}
\right)^2}{(1+d)^2}\\
\label{phasefk2}
{\cal G}(x) &=& - \frac {d \alpha \sin(\alpha x)}{1+d}\\
\label{phasefk3}
{\cal H}(x) &=& -\frac {d \alpha^2\cos(\alpha x)}{2(1+d)} \ \ . 
\label{phasefk4}
\EA

Setting the right hand side of Eq.~(\ref{nfk1}) to zero in order to
seek for stationary solutions, and thinking of $x$ as the time we can
view Eq.~(\ref{nfk1}) as a nonlinear evolution equation for a
one-dimensional field with a `time'-periodic parametric driving due to
the boundaries. Present knowledge on spatio-temporal chaos and pattern
formation can in principle be applied to analyze the behavior of this 
resulting evolution equation. General results are not abundant, 
however. In the next Section further approximations will be introduced
in order to facilitate the analysis and establish the existence of
stationary spatially-chaotic solutions. 

\section{SINGLE-TRANSVERSE-MODE APPROXIMATION}
\label{Low}

For definiteness, in the rest of the Paper we will consider just null
Dirichlet boundary conditions, that is the field $\psi$ takes the value
zero at the transverse boundaries: $\psi_0(x)=\psi_1(x)=0$. In this
case our model in the form of Eq.~(\ref{nfk1}) has the trivial
solution $\tilde\psi(x,\tilde y,t)=0$. The stability analysis of this
solution for the case of a rectangular domain of width $l$ leads to an
eigenvalue problem for the linearized equation. For $a>a_{c}=(\frac
{\pi}{l})^2$ the eigenfunctions factorize into longitudinal
($\exp{(ik_x x})$) and transverse ($\sin(k_{\tilde y} \tilde y)$)
modes: $\tilde \psi_{\lambda,\bf k}(x,\tilde y,t)=\exp(\lambda
t)\exp(ik_x x) \sin(k_{\tilde y} \tilde y)$, with $k_{\tilde y}=\frac
{\pi}{l}m, m=1,2,\ldots$, $k_x$ real, $\lambda$ satisfying the
dispersion relation $\lambda=a - k^2$, and $k^2=k_x^2+k_{\tilde y}^2$.
The unstable modes are then those satisfying the condition $k^2<a$. The
first unstable mode corresponds to $(k_x,k_{\tilde y})=(0,\frac
{\pi}{l})$, which becomes unstable at the critical value $a_c=(\frac
{\pi}{l})^2$. The transverse modes are discretized in multiples of
$\frac {\pi}{l}$ due to the boundary condition. If in addition, we
require $L$-periodicity in the longitudinal coordinate $x$, $k_x$ will
be also discrete, but provided that $L \gg l$ this discretization
will be much finer than the transverse one. The value of the parameter
$a$ controls how many modes are linearly unstable. If the transverse
size $l$ is small enough for the control parameter to satisfy the
condition $(\frac {\pi}{l})^2 < a <(\frac { 2 \pi}{l})^2$, there would
be just one linearly unstable transverse mode, with many associated 
longitudinal unstable modes. Close enough to the instability
threshold, we can try an approximate solution of the form $\tilde
\psi (x,\tilde y,t) = A(x,t) \sin(\pi \tilde y)$ and write an 
evolution equation for the amplitude $A(x,t)$ of the first transverse
mode. 

We are interested however in a domain which is not a rectangular strip
but a undulated channel. The coordinate change bringing our domain into
a rectangular one renders the variables non-separable and the linear
problem is no longer solvable analytically. However, for small
deviations from the uniform channel a perturbation scheme can be used.
In the same vein as in the preceding paragraph we try, for our
undulated domain, the ansatz  $\tilde\psi (x,\tilde y,t) = A(x,t)
\sin(\pi \tilde y) + {\cal O}(d_0,d_1)$, assuming that $a$ is
close to the threshold imposed by the small $l$, and that the size of
the channel undulations is small. We call this approach a 
single-transverse-mode approximation (STMA). 

Projecting Eq.~(\ref{nfk1}) onto the single transverse mode present in
the ansatz and neglecting higher order contributions we get the
following evolution equation for the amplitude $A$:
\BE
\partial_{t} A =
\partial^{2}_{xx} A + \beta(x) \partial_{x}A+\omega^{2}(x) A
-\frac {3}{4} A^{3}
\label{osc}
\EE
where
\BE
\beta(x)=\frac {\Delta_{x}}{\Delta}
\label{beta}
\EE
and
\BA
\omega ^{2}(x)=a &+& \frac {\Delta \Delta_{xx} - 2 \Delta_{x}^{2}}
{2\Delta^{2}} - \left(\frac {\pi}{\Delta}\right)^{2}
\left(1 + y_{1x} y _{ox} \right) \nonumber\\
&-& \left(2 \pi^{2} -3 \right) {\beta^{2}\over 6}
\label{omega}
\EA

We have checked that a more rigorous, but lengthy, approach based on a
multiple scale expansion leads to the same result \cite{Doelman95}. 
The stationary patterns satisfy the time independent version of 
Eq.~(\ref{osc}). In terms of the coordinate $x$ considered as a time, 
one can view this spatial dynamical system as a parametrically forced
nonlinear oscillator: domain undulations provide a periodic driving on
the frequency of the system $\omega ^{2}(x)$. In turn, $\beta(x)$ is a
``dissipation'' term which can be positive or negative depending on
$x$. This $x$-modulation of $\beta$ comes from the longitudinal
variation of the vertical width. The integral of the ``dissipation''
$\beta(x)$ on one period of the oscillating boundaries $T =
2\pi/\alpha$ is $\int_{x_0}^{x_0+T} \beta(x) dx = \ln
\frac{\Delta(x_0)}{\Delta(x_0+T)}=0$. This shows that although system
(\ref{osc}) is locally dissipative, it is effectively conservative
over one period of the modulation. This implies, in particular, that
the stroboscopic map associated with the system (\ref{osc}) is area
preserving. 

In the general case, Eq.~(\ref{osc}) can be simplified by removing the
dissipation term with the change: $A(x,t)=\exp\left(- \frac 
{1}{2} \int \beta(x) dx\right) \rho(x,t)= 
\rho(x,t)/\sqrt{\Delta(x)}$. The new equation reads: 
\BA 
\partial_{t} \rho & = &
\partial^{2}_{xx} \rho + \Omega^{2}(x) \rho -\frac {3}{4 \Delta} \rho^{3} 
\label{nosc}\\
\Omega ^{2} & = & a-\left(\frac {\pi}{\Delta}\right)^{2} \left(1 + 
y_{1x} y _{ox}\right) - \frac {\pi^{2} +3} {12} \beta^{2} \ \ .
\label{nomega}
\EA

A particular case occurs when the transverse distance between the two
channel borders does not vary along the longitudinal direction, so that the
dissipation term vanishes identically, i. e. $\beta (x)=0$. For the
sinusoidal channel this happens when $\phi=0$ and $d_0=d_1=d$
(Fig.~\ref{fdomain}b). In this case, the amplitude equation is reduced to 
\BE
\partial_{t} A =
\partial^{2}_{xx} A + \omega^{2}(x) A -\frac {3}{4} A^{3}
\label{phaseosc}
\EE
with
\BE
\omega ^{2} = a- \left(\frac {\pi}{1+d}\right)^2 - \frac 
{\left(\pi d \alpha\right)^2}{4 (1+d)^2}\sin^2 (\alpha x) \ \ . 
\label{phaseomega}
\EE
When $\partial_t A =0$, (\ref{phaseosc}) is a dissipative Mathieu 
equation modified by the addition of a cubic nonlinear term or, 
equivalently, a parametrically forced Duffing oscillator. This 
equation is known to have chaotic solutions \cite{Nayfeh95}. 

\bc
\begin{figure}[t]
\epsfig{file=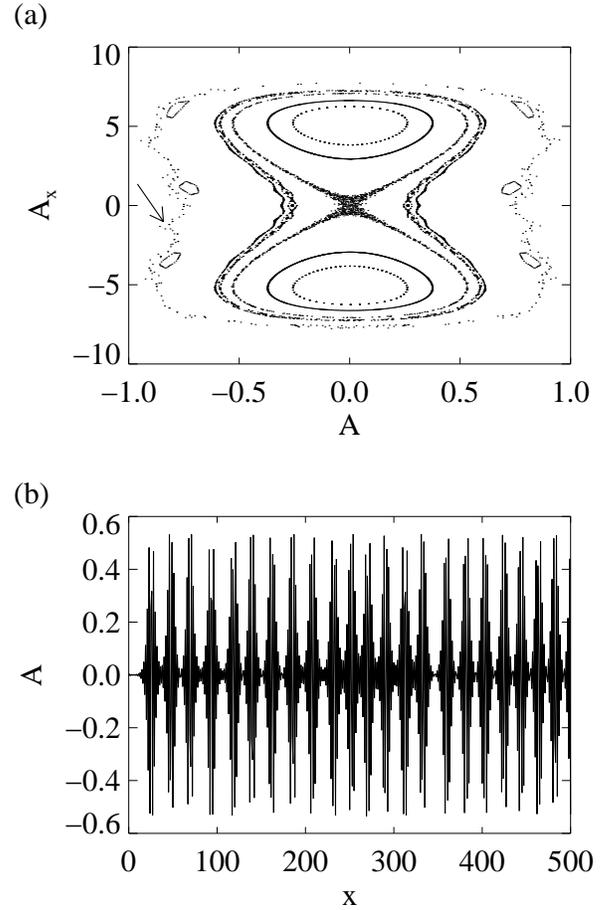,width=1.\linewidth}
\caption{ (a) Stroboscopic
Poincar\'e map of the phase space of the system (\protect \ref{duffing})
for the values $d_{0}=d_{1}=1$, $\alpha=2\pi$, $a=17$, $\phi=\pi$. KAM
tori and chaotic trajectories inbetween are clearly seen.
The arrow indicates the approximate
location of the fractal boundary separating bounded and unbounded
trajectories. (b) The
chaotic configuration corresponding to the cloud of points surrounding
the origin in the stroboscopic map. }
\label{fphasespace}
\end{figure}
\ec

The general time-independent case of Eq.~(\ref{osc}) [or 
Eq.~(\ref{nosc})] reads: 
\BE 
\partial_{xx}A = -\omega^2(x)A +\frac{3}{4} A^{3}- \beta(x)\partial_x A ~.  
\label{duffing}
\EE
In the absence of undulations ($d_0=d_1=0$) parametric forcing and 
dissipation vanish and the equation is both Hamiltonian and integrable. It
has a region in phase space $(A,A_x)$ close to the origin where motion is
bounded and regular. Beyond the separatrices of the two saddle points $(A,
A_x)=\pm 2/\sqrt{3} (a -\pi^2, 0)$, trajectories escape to infinity. When 
undulations are introduced, separatrices of the saddle points deform and may
cross. It is well known that for perturbed Hamiltonian systems, separatrix
intersections indicate the onset of chaos. In our system, in addition to
chaotic bounded trajectories, separatrix intersections lead also to
fractalization of the phase-space boundary dividing bounded and unbounded 
trajectories. Melnikov theory provide us with the tools to determine
analytically the necessary conditions for separatrix intersection and the
occurrence of chaos. Following \cite{Nayfeh95}, the Melnikov function
$M(\theta)$ can be calculated for small $d$. For example, in the case 
$\phi=\pi$, $d_0=d_1=d$ one finds: 
\BE
M(\theta)= f(a,\alpha) \sin(\alpha \theta) ~.
\label{melnikov}
\EE
The fact that the function $M(\theta)$ has zeros as a function of 
$\theta$ indicates that separatrix intersection and chaotic 
behavior occur. Chaotic behavior appears for arbitrarily small 
sizes of the channel undulations. We expect similar behavior to 
occur for other values of $\phi$, $d_0$ and $d_1$. 

\bc
\begin{figure}[t]
\epsfig{file=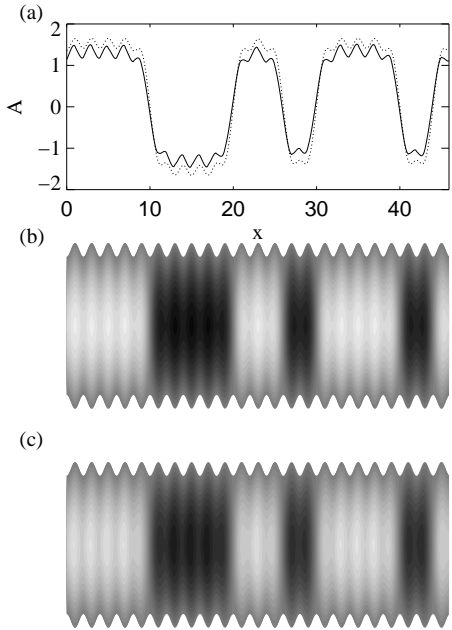,width=.7\linewidth,angle=0}
\vspace{1cm} \caption{ Stationary solution of the STMA~(\ref{osc}) 
compared to a fully two-dimensional simulation of (\ref{fk}) for 
the parameter values $d_0=d_1=0.1$, $\alpha=\pi$, $a=10$, 
$\phi=\pi$. a) Stationary solution $A(x)$ of the STMA (dotted) and 
the on-axis values of the actual two-dimensional stationary 
solution $\psi(x,y=0)$ (solid). b) The full two-dimensional 
solution of (\ref{fk}) represented on a gray-scale. White 
corresponds to the highest values of $\psi$ and black to the 
lowest. c) Reconstruction of the two-dimensional field from the 
STMA ($\psi(x,y)=A(x)\sin(\pi \tilde y)$) on the same gray-scale 
as in b). } 
\label{fcomparacion} 
\end{figure}
\ec

To illustrate the chaotic behavior of the stationary STMA 
(\ref{duffing}) we show in Fig.~(\ref{fphasespace}a) its numerical 
stroboscopic map. The dots are values of $(A,A_x)$ at multiples of 
the forcing period $T=2\pi/\alpha$ for a set of initial 
conditions. Several regions dominated by chaotic trajectories and 
separated by KAM tori (the closed curves corresponding to 
quasi-periodic solutions) are clearly recognized in the picture. 
Also, the approximate location of the fractal boundary separating 
bounded trajectories from those escaping to infinity is pointed by 
an arrow. Melnikov analysis also implies the existence of a dense 
set of unstable periodic orbits in the vicinity of the separatrix 
intersection on both sides of the fractal boundary mentioned 
before. These periodic orbits and the fact that they constitute a 
skeleton of the chaotic trajectories, are important for our 
analysis because the periodic boundary conditions in the 
$x$-direction select them out from the uncountable many other 
possible solutions of Eq.~(\ref{duffing}). 

\section{NUMERICAL STMA AND TWO-DIMENSIONAL TIME-INTEGRATIONS}
\label{Num}

In the previous section, we have shown both analitically and numerically that
simple undulated boundaries may induce spatially chaotic steady solutions in
our simple model (\ref{osc}). In this section, we discuss the accuracy and
range of validity of the STMA and the physical relevance of its solutions by
comparing with the numerical integration of the full model (\ref{fk}).

In the first place, we discuss the results of such integration for the
particular case $\phi=\pi$ and $d_0=d_1=d$. Starting from random
initial conditions the system, after a time long enough, settles in a
disordered stationary configuration, as shown in
Fig.~(\ref{fcomparacion}). There we present the stationary
configuration $A(x)\equiv A(x,t\rightarrow \infty)$ obtained by direct
integration of Eq.~(\ref{osc}). In the same plot, a longitudinal cross
section of the asymptotic field obtained from a simulation with the
full two-dimensional model is shown for comparison. We use as the
initial condition of the two-dimensional problem the solution from the
STMA for the same parameter values, $\psi=A(x)\sin(\pi\tilde y)$, to
provide an approximate stationary solution to Eq.~(\ref{fk}). After a
short time of adjustment, the system settles in a stationary state that
is very close to the initial approximation. The full two-dimensional
field and its reconstruction from the STMA are also shown in
Figs.~(\ref{fcomparacion}b) and (\ref{fcomparacion}c), respectively.
This figure reveals a strikingly accurate fit of the STMA  solution and
the complete field simulation, a strong indication of the validity of
the approximation as a tool for analysis. The maximum absolute error of
the approximate solution is of the same order of magnitude in both the
undulated channel and a rectangular-domain test-case. This suggests
that the error is mainly due to the truncation at the first linear
transverse mode, but not to the peculiarities of the curved
boundaries. 

The accuracy of the STMA, $A(x,t)\sin(\pi\tilde y)$, breaks down when
the channel width increases or when strongly non rectangular domains
are considered. Nevertheless, we have also performed direct simulations
of Eq.~(\ref{fk}) for this last case. An example of the typical
behavior is shown Fig.~(\ref{fnonrectangular}) for boundaries defined
by  $d_0=d_1=1.0$, $\alpha=\pi$, $a=20$, $\phi=\pi$. Notice that the
resulting stationary configuration displays the same qualitative
features of the STMA solutions: disordered distribution of kinks
randomly pinned at some of the narrows of the channel. This is an
evidence of the fact that undulating boundaries may also be the source
of stationary spatial chaos in a two-dimensional system (\ref{fk})
beyond the regime well described by a single transversal mode. 

\bc 
\begin{figure}[t] 
\epsfig{file=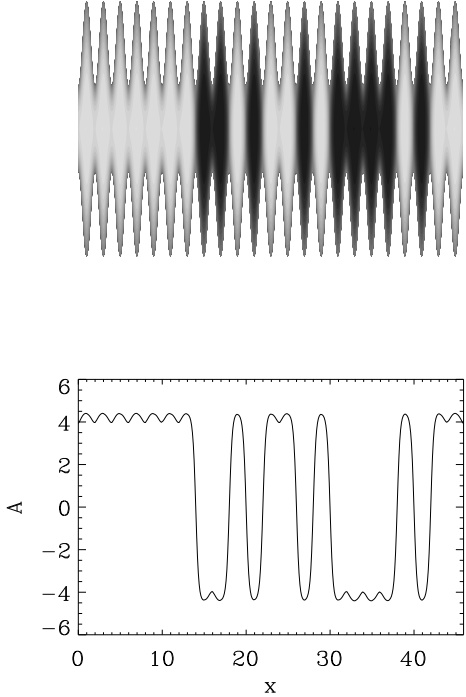,width=.85\linewidth} \caption{Two-dimensional 
steady state obtained by simulation of the system (\protect 
\ref{fk}) starting from random initial conditions. The amplitude 
of the field on the channel axis is shown in the lowest panel. 
Parameter values: $d=1.0$, $\alpha=\pi$, $a=20$, $\phi=\pi$.} 
\label{fnonrectangular} 
\end{figure} 
\ec

The results in Figs.~(\ref{fcomparacion}) and (\ref{fnonrectangular})
illustrate the physical mechanism behind the emergence of spatial chaos
in our  system. Let us  remember that Eq.~(\ref{fk}) evolves to
minimize the potential (\ref{V}). This minimization requires to reduce
as much as possible the length of the interfaces between $\psi_+$ and
$\psi_-$. Following this tendency, an interface that links the opossite
lateral banks of the channel and is far from any other interface will
evolve to lock into one of the narrows of the channel, where it is
shorter than in any other position. Detaching the interface from the
bank of the channel would imply a temporary increase of the potential
$V$ due to the necesary proliferation of new interfaces. Such a
potential increase is not allowed by the dynamics. Hence, the random
occupation (arising from random initial conditions) of the narrows of
the channel by kinks and anti-kinks finally builds up a spatially
chaotic stationary configuration. This argument, based on kink and
interface dynamics, clearly applies beyond the range of validity of the
STMA, where we have analytically established the existence of spatial
chaos, as evidenced in Fig.~(\ref{fnonrectangular}). With this
mechanism in mind, we can also conclude  that not all the chaotic
trajectories presented in Fig.~(\ref{fphasespace}) will lead to spatial
chaos stable in time: only those corresponding to an energetically
favorable (a local minima of the potential $V$) distribution of kinks
will be reached under time evolution. In particular the trajectory
plotted in Fig.~(\ref{fphasespace}b) corresponds to a temporally
unstable configuration. 

\bc 
\begin{figure}[t] 
\epsfig{file=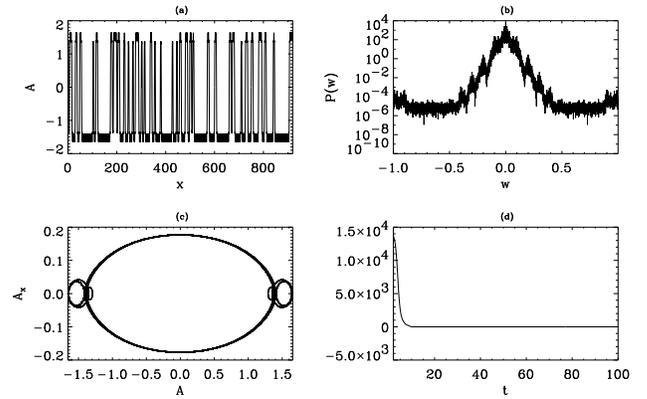,height=1.\linewidth,angle=90} \caption{ 
One-dimensional solution of the system (\ref{osc}) showing a 
steady chaotic configuration, its spatial power spectrum, the trajectory
in the 
projected phase space $(A,A_x)$, and the time evolution of the 
Lyapunov functional, showing that this is an attracting 
configuration. Parameter values as in Fig.~(\ref{fcomparacion}), 
but for a longer system} 
\label{freal} 
\end{figure}
\ec

Let us now try to get further insight about the chaotic nature of
the irregular spatial structures described above. Periodic 
boundary conditions in the longitudinal direction always force 
the system to converge not to a chaotic spatial configuration, but 
to a periodic one wich we have shown it may very well be of the 
maximal period. To justify the use of the chaotic qualifier we 
need to show that as the size of the system increases, these 
periodic configurations approach one that could be characterized 
as chaotic in some way. Of course, to numerically carry out this 
process we would need to consider very long channels. 
Unfortunately though, performing direct simulations on the fully 
two-dimensional model soon becomes computationally prohibitive as 
the number of the channel undulations increases. However, having 
demonstrated that the STMA accurately describes the qualitative 
features of the full model, we can concentrate our attention on 
the behavior of the approximate model Eq.~(\ref{osc}). 
 
In Fig.~(\ref{freal}) we summarize the results from the numerical 
integration of Eq.~(\ref{osc}) using the same parameter values as 
in Fig.~(\ref{fcomparacion}) with the exception of the domain size 
which now is much larger. An asymptotically stable configuration 
is shown in a) while c) displays the projection of the trajectory 
in phase space $(A,A_x)$. The power spectrum of $A(x)$ plotted in 
Fig.~(\ref{freal}b) shows the typical broadband feature 
characteristic of chaotic trajectories. As a further indication of 
the (aproximately) chaotic nature of this configuration, we have 
shown in Fig.~(\ref{fpoincare}) the stroboscopic map constructed 
from the trajectory in Fig.~(\ref{freal}). taking phase space 
points at `times' integer multiples of $T$. This map reveals an 
incipient self-similar fractal structure, also common in chaotic 
trajectories. All these facts together give compealing evidence 
that the trajectory, although $L$-periodic by construction, 
develops chaotic features as system size increases. 

\bc
\begin{figure}[t] 
\hspace{-1cm}
\epsfig{file=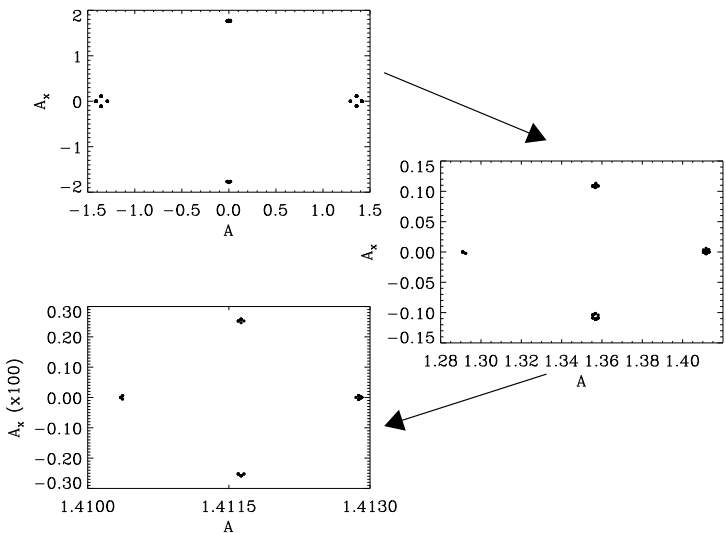,width=.8\linewidth,angle=90}
\vspace{.5cm}
\caption{ 
Stroboscopic map of the trajectory shown in Fig.~(\ref{freal}). 
Two succesive amplifications are shown, displaying
a self-similar structure typical of a fractal.}
\label{fpoincare} 
\end{figure}
\ec

Finally, as an illustrative measure of the asymptotic stability of this
solution, Fig.~(\ref{freal}d) displays the value of the Lyapunov
potential $V[\psi]$ evaluated along the time evolution of the field
$\tilde \psi(x,\tilde y,t)=A(x,t)\sin(\pi\tilde y)$. The functional
decreases in time, confirming the consistency of the STMA with the
exact dynamics of Eq.~(\ref{relax}), and the potential asymptotically
approaches a constant  value indicating that the field has reached a
local minimum of $V$. 

\section{CONCLUSIONS}

We have given evidence of the existence of stationary, stable, 
longitudinally chaotic spatial configurations induced by undulated 
boundaries in a simple two-dimensional reaction-diffusion model 
that does not otherwise display any kind of chaos. We have 
demonstrated that these type of boundaries can be convincingly 
mapped into spatially-periodic parametric modulations in a 
one-dimensional approximation to the original system. In a 
dynamical systems approach to the study of stationary solutions, 
these modulations play the role of a temporal time-periodic 
forcing capable to drive a nonlinear second-order ODE into chaotic 
behavior. The diffusive character of our original model ensures 
precisely that the relevant ODE is in fact, second order and that 
the presence of chaotic stationary solutions is expected at this 
level of approximation. Previous results 
exist\cite{Coullet87,Gorshkov94,Malkov95} showing that spatial 
periodic modulation of some parameters intrinsic to the dynamics 
may originate spatial chaos in relatively simple reaction 
diffusion models. However, to our knowledge, the present is the 
first example in which an straightforward dynamical systems 
approach is used to establish the existence of disorder in 
two-dimensional systems due to the influence of the boundaries. 

The consequence of our analysis is that chaotic configurations 
should {\sl exist} in virtually any of the experimental systems 
commonly used to study pattern formation, provided that boundary 
conditions such as those studied here are imposed. The {\sl 
stability} of these chaotic configurations should be discussed in 
each particular case. While in our model stability comes from the 
tendency of the dynamics to minimize (\ref{V}) therefore 
minimizing interface lengths and leading to pinning of these 
interfaces to the narrows of the channel, the mechanism for 
stability in other systems may be different. Apart from the direct 
application to pattern forming systems, the idea of 
boundary-generated spatial chaos could be speculatively 
transferred to other nonlinear extended dynamical systems of 
interest. For example, it is possible that low Reynolds number 
fluid flows through a space-periodically perturbed pipeline or 
even through a realistic channel of shape similar to the ones 
considered here can display frozen spatial chaos \cite{LeGuer}. A 
numerical search for this manifestation of ``frozen turbulence'' 
at the level of Navier-Stokes equations is currently in progress. 
We expect this observation to promote also experimental work both 
in the area of pattern formation and in hydrodynamics. 

\acknowledgements
 Financial support from PB94-1167 and PB97-0141-C02-01 is greatly 
acknowledged. 



\end{twocolumns}
\end{document}